\documentclass[10pt]{article}
\usepackage[english]{babel}
\usepackage{amsfonts}
\usepackage[dvips]{graphicx}

\begin{document}

\title{\bf On the notion of laminar and weakly turbulent elementary fluid flows: a simple mathematical model}

\date{August 2006}

\author{\bf Gianluca Argentini \\
\normalsize gianluca.argentini@riellogroup.com \\
\textit{Research \& Development Department}\\
\textit{Riello Burners}, 37048 San Pietro di Legnago (Verona), Italy}

\maketitle

\begin{abstract}
An elementary analytical fluid flow is composed by a geometric domain, a list of analytical constraints and by the function which depends on the physical properties, as Reynolds number, of the considered fluid. For this object, notions of laminar or weakly turbulent behavior are described using a simple mathematical model. 
\end{abstract}

\section{Elementary analytical fluid flows}

Let $L$ a length unit and $\bar{\Omega}$ a closed domain of $\mathbb{R}^n$, $n \geq 2$, formed by a small number (e.g. 2 or 3) of $L^2$ squares or $L^3$ cubes. Here we consider the plane case, but the notions are easly extensible to greater dimensions. Let $F$ a fluid entering in $\Omega$ from a single edge of a square and flowing out from another single edge.
Let $s$, $0 \leq s \leq 1$, a parameter which describes the position of a single fluid particle along the inflow edge, so that $s_0L$ is the initial position of the generic particle associated to the particular value $s_0$ of the parameter.
Let now $s \mapsto R(s)$ a function $R:[0,1] \rightarrow [0,1]$ which maps a value of the parameter $s$ to the value identifying the position reached by the particle on the outflow edge, so that this position is identified by the value $R(s)L$. If ${x,y}$ is a cartesian coordinates system in the plane of $\Omega$, assume that each streamline, or each particle path in lagrangian view, is described by a parametric curve $t \mapsto \Phi(t)=\left(x(t),y(t)\right)$, with $0 \leq t \leq 1$. The parameter $t$ could not be, in general, the time variable of the flow. Let this parametric representation be determined by a set $A$ of analytical conditions regarding $\Phi(t)$ and $\dot{\Phi}(t)$, that is the passage of the particle in some suitable points of $\Omega$ and the velocity field direction in some other (or the same) points.\\
An {\it elementary analytical flow} is a particular list $\mathbb{F}=\left\{\Omega,R,A\right\}$. We propose a simple mathematical model, at least for a particular example of domain $\Omega$, and show that the function $R(s)$ can identify the physical properties of the particular fluid $F$ flowing in the domain. For simplicity, in this work assume that the flow is steady.

\section{A mathematical model}

In this section let $\Omega$ the domain formed by three $L^2$ squares, the first two along the $x$-axis of a cartesian coordinates system from $0$ to $2L$, and the third above the second, from $L$ to $2L$ $y$-coordinates.

\begin{figure}[ht]
	\begin{center}
	\includegraphics[width=6cm]{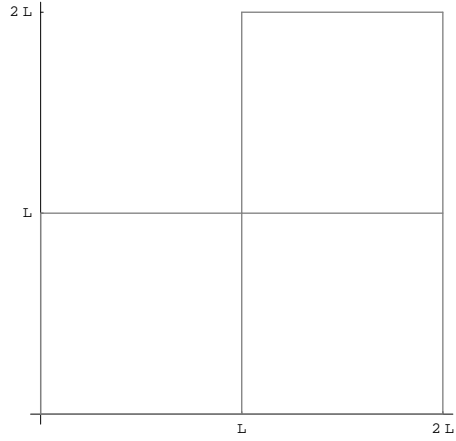}
	\caption{\small{\it Domain with three unit squares.}} 
	\end{center}
\end{figure}

\noindent The inflow edge is the segment $[(0,0),(0,L)]$ of the first square and the outflow edge is the segment $[(L,2L),(2L,2L)]$ of the third square.\\

\noindent Now the set $A$ af analytical conditions is so defined:\\
\indent $P_1$. at the inflow edge, for $t=0$ and position parameter $s$, let $\Phi(0)=(0,sL)$;\\
\indent $P_2$. at the outflow edge, for $t=1$ and position parameter $R=R(s)$, let $\Phi(1)=(L(1+R),2L)$;\\
\indent $P_3$. for $t=\frac{1}{2}$ a particle path intersects the diagonal line of the second square, that is the line of cartesian equation $y=-x+2L$; if a parameter $p$, $0 \leq p \leq 1$, describes the positions along this line, the condition is $\Phi(\frac{1}{2})=(L(2-p),pL)$;\\
\indent $D_1$. at the inflow edge the velocity is parallel to $x$-axis, so that for every $s$ $\dot{y}(0)=0$;\\
\indent $D_2$. at the outflow edge the velocity is parallel to $y$-axis, so that for every $s$ $\dot{x}(1)=0$.\\

\noindent Note that condition $P_3$ is a logical connection between $x$ and $y$ coordinates, so that for each component of $\Phi(t)$ the independent conditions are four: assuming, for semplicity, that the components are polinomial expressions on variable $t$, the candidates to satisfy the set $A=\{P_1,P_2,P_3,D_1,D_2\}$ of conditions are the cubics $x(t)=at^3+bt^2+ct+d$ and $y(t)=et^3+ft^2+gt+h$.\\

\noindent Let proceed with the computation of these cubics.\\
From derivation condition $D_1$ follows $g=0$. From passage condition $P_1$ follows $d=0$ and $h=sL$. From condition $D_2$ follows $c=-3a-2b$. With these partial results, from condition $P_3$ follows $b=-L(1+R)-2a$ and $c=a+2L(1+R)$. Now, using condition $P_2$, the computation gives $a=L(10-6R-8p)$, so that the first component is

\begin{equation}\label{xcomp}
	x(t)=L(10-6R-8p)t^3+L(-21+11R+16p)t^2+L(12-4R-8p)t
\end{equation}

\noindent With a similar computation, the second component is

\begin{equation}\label{ycomp}
	y(t)=L(4+6s-8p)t^3+L(-2-7s+8p)t^2+Ls
\end{equation}

\noindent Note that, with this model, the second component doesn't depend on function $R(s)$.\\
For semplicity, let $p=s$, that is the particles paths are not {\it disturbed} until the line $y=-x+2L$. In this case the $y$-component has the form

\begin{equation}\label{ycomp2}
	y(t)=L(4-2s)t^3+L(-2+s)t^2+Ls
\end{equation}
 
\section{The laminar case}

In the laminar case assume that particles paths don't intersect themselves, that is the outflow positions on the final edge of $\bar{\Omega}$ are the same as inflow positions. The analytical expression of this geometrical condition is simply

\begin{equation}\label{laminar}
	R(s)=1-s
\end{equation}

\noindent The laminar elementary analytical flow $\mathbb{F}=\{\Omega,A,R\}$ is so completely defined; its particles paths have the form

\begin{equation}\label{pathsLaminar}
	\Phi(t)=\{L(2-s)(2t^3-5t^2+4t),L(2-s)(2t^3-t^2)+Ls\}
\end{equation}

\noindent Note that the expression (\ref{laminar}) for $R(s)$ depends on the shape of the domain $\bar{\Omega}$, as expected; e.g., in the case of a rectangular domain with the outflow edge parallel to the inflow edge, the expression should be $R(s)=s$.

\begin{figure}[ht]\label{laminarFigure}
	\begin{center}
	\includegraphics[width=6cm]{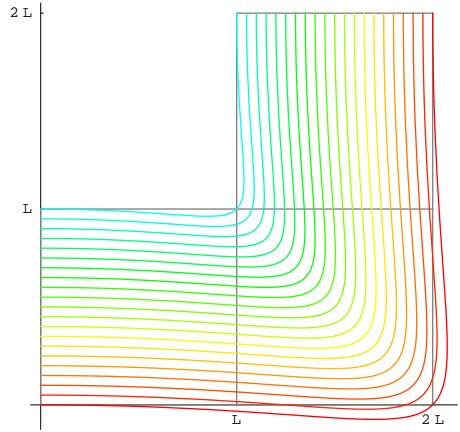}
	\caption{\small{\it Particles paths for laminar elementary flow.}} 
	\end{center}
\end{figure}

\section{The weakly turbulent case}

The weak turbulence notion (\cite{ruelle}) regards the physical situation of a flow where turbulence is not fully developed and it could be described by deterministic mathematical law (\cite{roux}). For an elementary analytical flow, such mathematical description can be made by a suitable choice of the function $R(s)$, which maps the initial inflow position of a particle to its outflow position.\\
In order to find a possible expression for $R(s)$, assume that some geometrical point of the final edge can be reached by more than one fluid particle, so that paths intersects themselves, as expected in a turbulent flow. Also, the analytical expression of $R(s)$ should be such that, in the case of suitable conditions, the laminar case (\ref{laminar}) can be a mathematical limit for great values of viscosity ({\it viscous limit}, see \cite{constantin}). A mathematical condition is that $R(s) \in [0,1]$ for every $s$. Assume also that weak turbulence, as expected, is associate to the analytical condition $R \in C^m([0,1])$ for some integer $m \geq 0$.\\
Note that a possible expression satisfying all the previous conditions is $R(s)=\frac{1}{2}\left[1+sin(\alpha)\right]$, where $\alpha=\alpha(s)$ could be computed using the viscous limit. Assume that this limit is defined for Reynolds number $\mathbb{R}e$ approaching a value $c_{\mathbb{F}}$, $\mathbb{R}e \rightarrow c_{\mathbb{F}}$, with $c_{\mathbb{F}}$ a constant which should depend on the nature of the fluid and on the geometrical and physical properties of the flow. The most simple form for $\alpha$ is $\alpha=a\hspace{0.1cm}\mathbb{R}e\hspace{0.1cm}s+b$, therefore

\begin{equation}
	\mathbb{R}e \rightarrow c_{\mathbb{F}} \Rightarrow \alpha \rightarrow ac_{\mathbb{F}}s+b
\end{equation}

\noindent Assuming $ac_{\mathbb{F}}s+b$ sufficiently small, for $\mathbb{R}e \approx c_{\mathbb{F}}$ follows $sin(\alpha) \approx ac_{\mathbb{F}}s+b$, therefore it must be

\begin{equation}
	R(s) = \frac{1}{2}\left[1+ac_{\mathbb{F}}s+b\right] = 1-s
\end{equation}

\noindent from which $a=-\frac{2}{c_{\mathbb{F}}}$ and $b=1$. So the function describing a weak turbulent elementary flow is

\begin{equation}\label{weakTurbulent}
	R(s) = \frac{1}{2}\left[1+sin\left(1-\frac{2}{c_{\mathbb{F}}}\hspace{0.1cm}\mathbb{R}e\hspace{0.1cm}s\right)\right]
\end{equation}

\noindent Note that for high Reynolds number the function $R(s)$ {\it covers} the interval $[0,1]$ many times, so that the degree of turbulence increases.

\begin{figure}[ht]\label{viscousFigure}
	\begin{center}
	\includegraphics[width=7cm]{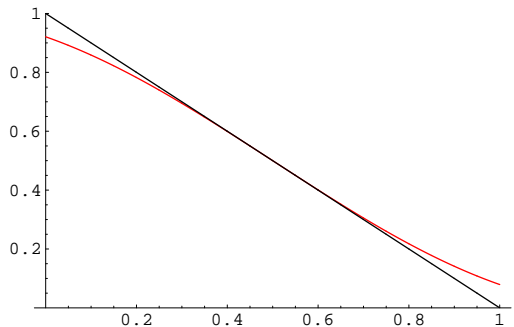}
	\caption{\small{\it Viscous limit (red) $R(s)=\frac{1}{2}\left[1+sin\left(1-2s\right)\right]$, compared to laminar case $R(s)=1-s$.}} 
	\end{center}
\end{figure}

\begin{figure}[ht]\label{turbulentFigure}
	\begin{center}
	\includegraphics[width=11cm]{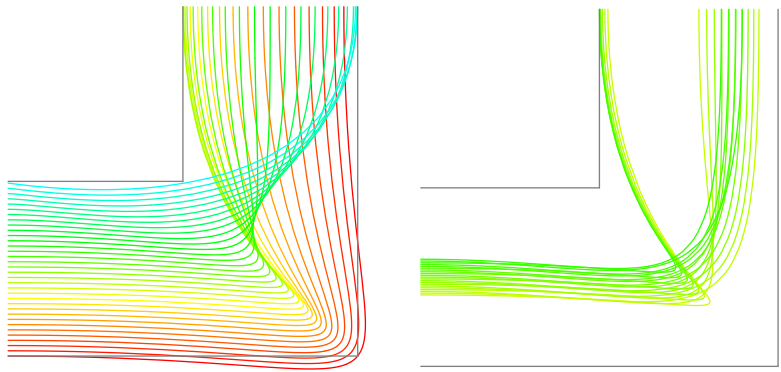}
	\caption{\small{\it Weakly turbulent elementary flows: at left, particle paths for $\frac{\mathbb{R}e}{c_{\mathbb{F}}}=3$; at right, some particle paths in the case $\frac{\mathbb{R}e}{c_{\mathbb{F}}}=2\hspace{0.1cm}10^4$.}} 
	\end{center}
\end{figure}

\section{Concluding remarks and further developments}

The notion of elementary flow is characterized by a domain, a list of analytical constraints and by the function $R=R(s):[0,1] \rightarrow [0,1]$ which depends on the geometrical and physical properties of the considered fluid. A particular weakly turbulent behavior can be obtained when $R(s)$ is a periodic function. Then laminar behavior is a viscous limit of this model.\\
Some questions arise.\\
It could be investigated the relation between particles paths (or streamlines) $\Phi(t)$ and the solutions of Navier-Stokes equations, probably obtaining a mathematical relation between $R$ and pressure, and an expression for the flow constant $c_{\mathbb{F}}$.\\
In the case of evolution, e.g. if the $t$ parameter is proportional to time, function $R(s)$ could be characterized using the Richardson's Law (\cite{boffetta}) on particles dispersion in turbulent flows.\\
Can be full developed turbulence a limit of this weak turbulence model? For very high Reynolds numbers the function $R(s)$ expressed by (\ref{weakTurbulent}) formula has no limit for every $s$, or has not a deterministic behavior: the vertical line $[0,1]$ is an accumulation line for its graph, therefore no prediction about the outflow position of a particle is possible. Perhaps it can be useful a convergence, for some kind of metric, of a succession $\{R_n(s)\}_{n=1,2,3,...}$ of continuous $R$s to some step function, in order to describe full turbulence flow and its phenomemon of intermittency (\cite{meneveau}).\\
Finally, can be a real fluid flow described by some kind of combination of elementary flows? Perhaps a combination, or a succession, of elementary flows with a common function $R(s)$ and variable analytical constraints $\{A_n\}_{n=1,2,3,...}$ could be a useful representation of a real flow of a single fluid.

\end{document}